\documentclass[sigconf, authorversion]{acmart}
\usepackage[english]{babel}
\usepackage{graphicx}
\usepackage{epstopdf}
\usepackage{acronym}
\usepackage{balance}
\usepackage{amsmath}
\usepackage{mathtools}
\usepackage{comment}
\usepackage{caption}
\usepackage{enumitem}
\usepackage{adjustbox}
\usepackage{acronym}
\usepackage{array}
\usepackage{makecell}
\usepackage{pifont}

\acrodef{GNSS}{Global Navigation Satellite Systems}
\acrodef{SDR}{Software Defined Radio}
\acrodef{IRA}{IRIDIUM Ring Alert}
\acrodef{LEO}{Low-Earth Orbit}
\acrodef{IoT}{Internet of Things}
\acrodef{STL}{Secure Time and Location}

\title{GNSS Spoofing Detection via Opportunistic IRIDIUM Signals}

\author{Gabriele Oligeri, Savio Sciancalepore, Roberto Di Pietro}
\affiliation{
  \institution{Division of Information and Computing Technology \\ College of Science and Engineering, Hamad Bin Khalifa University}
  \state{Doha, Qatar}
}

\acmYear{2020}\copyrightyear{2020}
\setcopyright{acmcopyright}
\acmConference[WiSec '20]{13th ACM Conference on Security and Privacy in Wireless and Mobile Networks}{July 8--10, 2020}{Linz (Virtual Event), Austria}
\acmBooktitle{13th ACM Conference on Security and Privacy in Wireless and Mobile Networks (WiSec '20), July 8--10, 2020, Linz (Virtual Event), Austria}
\acmPrice{15.00}
\acmDOI{10.1145/3395351.3399350}
\acmISBN{978-1-4503-8006-5/20/07}

\newcolumntype{P}[1]{>{\centering\arraybackslash}p{#1}}
\newcommand{\cmark}{\ding{51}}%
\newcommand{\xmark}{\ding{55}}%

\graphicspath{{figures/}}

\begin{document}

\begin{abstract}
    In this paper, we study the privately-own IRIDIUM satellite constellation, to provide a location service that is independent of the GNSS. In particular, we apply our findings to propose a new GNSS spoofing detection solution, exploiting unencrypted IRIDIUM Ring Alert (IRA) messages that are broadcast by IRIDIUM satellites. 
    
    To achieve the above-introduced objective, we firstly reverse-engineer many parameters of the IRIDIUM satellite constellation, such as the satellites speed, packet interarrival times, maximum satellite coverage, satellite pass duration, and the satellite beam constellation, to name a few. Later, we adopt the aforementioned statistics to create a detailed model of the satellite network. Subsequently, we propose a solution to detect unintended deviations of a target user from his path, due to GNSS spoofing attacks. We show that our solution can be used efficiently and effectively to verify the position estimated from standard GNSS satellite constellation, and we provide constraints and parameters to fit several application scenarios. All the results reported in this paper, while showing the quality and viability of our proposal, are supported by real data. In particular, we have collected and analyzed hundreds of thousands of IRA messages, thanks to a measurement campaign lasting several days. All the collected data ($1000+$ hours) have been made available to the research community.
    
    Our solution is particularly suitable for unattended scenarios such as deserts, rural areas, or open seas, where standard spoofing detection techniques resorting to crowd-sourcing cannot be used due to deployment limitations. Moreover, contrary to competing solutions, our approach does not resort to physical-layer information, dedicated hardware, or multiple receiving stations, while exploiting only a single receiving antenna and publicly-available IRIDIUM transmissions. Finally, novel research directions are also highlighted.
        
\end{abstract}




\maketitle

\section{Introduction}
\label{sec:intro}

\ac{GNSS} technologies today are used massively in several application scenarios, spanning from turn-by-turn terrestrial navigation, airborne and maritime navigation, timing purposes in smart grid, to name a few~\cite{yasuda2020_csur}. Despite Global Positioning System (GPS) by the US is the most famous, in the last years several region-specific solutions with the same aim have appeared, such as the European GALILEO, the Russian GLONASS, and the Chinese BEIDOU, to name a few. The vast majority of the functionalities provided by GNSS systems are open and available to the public, e.g., GPS provides an unencrypted (and un-authenticated) signal. While the lack of encryption fueled their adoption worldwide, in many use-case scenarios, unfortunately, the lack of authentication makes these systems prone to cybersecurity attacks. Indeed, the adversary can easily generate fake GNSS signals, and broadcast them to a set of targets, biasing the computation of their actual positions. We refer to the previous behavior as \emph{spoofing}~\cite{schmidt2016survey}.

While requiring minimum background knowledge, cheap hardware, and free-to-use software, location spoofing is easy to play, but hard to detect and mitigate. Indeed, since GNSS satellites are orbiting in a Medium Earth Orbit (MEO), far away from the ground (about 20,000 Km), GNSS signals reach the ground with a very weak signal-to-noise ratio. Therefore, an adversary can easily overcome the aforementioned received signal strength, being always much closer to the target device than the satellite constellation. Eventually, the (spoofed) signals transmitted by the adversary reach the target device and---having a higher SNR---are considered as the actual ones, and adopted for the final computation of the device's location~\cite{Tippenhauer2011}. The low strength of the GNSS received signal makes GNSS technologies also vulnerable to jamming, being it intentional (attack) or unintentional. While there is an active research community on localization of malicious jamming sources and mitigation of their effect, the unintentional jamming, due to radar, radios, and electromagnetic interference (e.g. from pumps and engines), is more difficult to mitigate, and can cause significant issues to the correct reception of GNSS satellite signals, by breaking the link for even long period of time~\cite{mpitz2009}.

During the years, several techniques have been adopted to detect and mitigate GNSS spoofing attacks~\cite{schmidt2016survey}. Some of them analyze the physical layer, trying to distinguish if the signals are coming from one only source (that could be a potential spoofer) or a set of sources, such as a genuine satellite constellation~\cite{magiera2015_jart}. Other solutions resort to crowd-sourcing, i.e., combining context information from neighbors, ad-hoc infrastructure, and cellular network information. However, the current state of the art does not take into account the specific scenario where signal sources are, by design, solo sources, or when just a few additional sources can be leveraged, but for a limited, intermittent, not predictable, period of time. 
In such scenarios, the deployment of additional infrastructures might be impractical, and the usage of physical layer information might not be possible due to either the lack of the underlying physical phenomenon, or because of the high price of ad-hoc hardware needed to exploit it.

A striking example is represented by open seas navigation, where ships resort only to GNSS positioning for computing their current location and heading up to the destination. Indeed, open seas navigation is witnessing an emergent trend, and similarly to the automotive industry, ships are becoming more and more autonomous. According to recent forecasts, a landscape of fully autonomous vessels is predicted by the year 2035, where GNSS technology will be massively adopted to provide location awareness~\cite{septentrio}. In such a scenario, GNSS jamming and spoofing will become dreadful threats, since autonomous vessels---even more than the human-controlled ones---require a trusted and precise localization procedure~\cite{caprolu2020_commag}.

{\bf Contribution.} In this paper, we provide two main contributions. First, we reverse-engineer several infrastructure configuration parameters of the IRIDIUM constellation. Second, leveraging the previously mentioned result, we exploit opportunistic \ac{IRA} messages to provide yet another independent localization solution.
As for the former contribution, one should notice that, though \acl{IRA} are accessible, all other information about IRIDIUM are treated as proprietary by the infrastructure owner: Iridium Communications Inc.. To overcome this lack of information, we ran an extensive measurement campaign using commercially available \acp{SDR}, acquiring \acl{IRA} messages for more than $1000$~hours. Using this large dataset, we reverse-engineered several network configuration parameters of the IRIDIUM constellation, including the satellites speed, packet interarrival times, maximum satellite coverage, satellite pass duration, and the satellite beam constellation---all these data have been released as open-source~\cite{crilab}. We used the above parameters to model the IRIDIUM constellation. Later, we moved to exploit opportunistic \ac{IRA} messages, combining them with the information acquired during the reverse-engineering process,  to estimate the current position of the receiving device.

Our research contributes not only to validate the GNSS information, but also to provide yet another independent localization solution---although being slightly inaccurate compared to the GNSS, given the harsh system conditions---in the presence of GNSS jamming. Being the IRIDIUM constellation characterized by worldwide availability, our solution can be used even in unattended scenarios (deserts and oceans), where no other ad-hoc in-land network infrastructure can be used for crowd-sourcing, and where the available GNSS hardware cannot be replaced or updated with multiple-antenna setups. Finally, we also highlight some novel, challenging research directions.

{\bf Paper organization.} The paper is organized as follows. Section~\ref{sec:related} reviews recent related work on GNSS spoofing detection and mitigation, Section~\ref{sec:iridium} introduces the main system features of the IRIDIUM satellite platform, Section~\ref{sec:adversary_model} depicts the adversarial model assumed throughout the work, Section~\ref{sec:measurement_setup} illustrates our measurement setup, Section~\ref{sec:iridium_data_analytics} includes the features of the IRIDIUM constellation platform inferred through our measurement campaign, Section~\ref{sec:position_error_estimation} provides the details of our spoofing detection technique based on opportunistic \ac{IRA} messages, while Section~\ref{sec:comparison} provides a systemic comparison between our proposal and the most important related work. Finally, Section~\ref{sec:conclusion} draws conclusions and highlights our future research directions.

\section{Related Work}
\label{sec:related}

\ac{GNSS} spoofing identification, detection, and mitigation are all research areas that attracted the attention of a lot of researchers during the years. Major contributions are coming from the idea of resorting to other over-the-air information to cross-check the actual position of the targeted spoofed device.

A very recent contribution~\cite{oligeri2019_wisec} involves crowd-sourced information from cellular network access points to verify the actual position of the device. Authors showed that beacons broadcast from cellular networks access points can be exploited to retrieve their position, and therefore, checking the consistency of the current position of the device received from the GPS constellation.

Authors in~\cite{jansen2016_acsac} showed that multiple (colluding) GPS receivers can be used to detect a spoofer. Authors showed that by leveraging spatial noise correlations, the false acceptance rate of the countermeasure can be improved while preserving the sensitivity to attacks. 
A collaborative detection scheme is proposed in~\cite{milaat2018_commlett}. The authors assume a set of vehicles exchange measured GPS code pseudo-ranges using a dedicated short-range communication. Each vehicle elaborates on the exchanged GPS data and derives independent statistics. These statistics are then exploited to detect high correlations in the time of arrival of spoofed GPS signals, and therefore, highlighting the presence of a GPS spoofer.

A physical layer solution is proposed by~\cite{zhang2012_milcom}, involving two antennas with different radiation patterns. Genuine GPS signals are supposed to come from several sources (satellites), and therefore, exploiting a two-antenna deployment involves a different signal-to-noise ratio measured by the two antennas. If the two antennas experience the same signal-to-noise, it means the signal is coming from the same source, and therefore, a GPS spoofer is likely to be in the neighborhood.

Authors in~\cite{sciancalepore2018_cns} proposed a solution exploiting meteor burst communications to verify the location of the spoofed device. The solution exploits the radio reflection properties of ephemeral meteor trails, combined with multiple anchors that can be deployed even hundreds of Km from the receiver. Although being robust to jamming and eavesdropping, their solution requires dedicated infrastructure.
Another crowd-sourcing solution has been proposed by~\cite{Jansen2018}. The authors proposed the deployment of multiple nodes at the ground to monitor the air traffic from GPS-derived position advertisements that aircraft periodically broadcast for air traffic control purposes. Spoofing attacks are detected and localized by the independent infrastructure on the ground which continuously analyzes the contents and the times of arrival of these advertisements. 

A vision-based GPS spoofing detection method for unmanned aerial vehicles (drones) is proposed by~\cite{qiao2017_cis}. The on-board camera and the inertial measurement unit are used to get the velocity and position of the aircraft, thus detecting unexpected changes in the flight path.
Another cooperative solution has been proposed by~\cite{heng2015_tis}. The GPS signal is verified by resorting to a network of cooperative GPS receivers. Each receiver in the network correlates its version of the signal with those received by other receivers to detect spoofing attacks.
The authors in~\cite{magiera2015_jart} exploited phase delay measurements from an antenna array to infer between single and multiple GPS sources, and therefore, detect the presence of a spoofer.

In Section~\ref{sec:comparison} we will provide a comparison between our solution and the ones introduced in this section, based on some distinctive system requirements.
We remark that the Iridium satellite network has been previously exploited for localization purposes by~\cite{leng}. The authors exploited physical layer information and a reference anchor in a well-known position to locate the receiver by resorting to both time difference of arrival (TDOA) and frequency difference of arrival (FDOA) techniques. Another similar contribution is provided by~\cite{Tan2019_access} and~\cite{tan2020_electronics}. The authors exploited the instantaneous Doppler effect of \ac{IRA} messages to infer the position of the receiver. However, using the Doppler shift for GNSS spoofing detection would require the deployment of multiple receivers, organized in ad-hoc network infrastructure, typically not available in deserts and oceans~\cite{ghose2015}, \cite{schafer2016}.

Finally, we recall that localization and spoofing detection are two different research problems, characterized by different system requirements. Localization techniques are used to either replace or support the traditional GNSS signals, while spoofing detection schemes are usually designed to work in conjunction with the GNSS, taking an active part (raising alarms and indicating corrections) only when they detect inconsistencies between the position reported by the GNSS and the one obtained via local computations. 

\section{Background on IRIDIUM}
\label{sec:iridium}

The IRIDIUM satellite platform has been described for the first time in the seminal paper~\cite{iridium}. The platform has been set up in 1993, and it is mainly constituted by a set of \ac{LEO} satellites, able to guarantee the full earth coverage for data and voice communications. IRIDIUM RF operations take place in the L-band, i.e., in the range $[1,616 - 1,626.5]$~MHz, by resorting to dedicated satellite-phones, while the communications between the satellites happen in the 23GHz band. The name of the satellite platform is due to the initial intended number of satellites, being 77 the atomic number of the chemical element Iridium. However, the final deployment of the satellite constellation counts only 66 satellites, which are the ones necessary to guarantee coverage for the whole Earth's surface. 
On the ground, to initiate and receive IRIDIUM calls, a user can leverage dedicated satellite phones and pagers provided by companies such as Motorola and Kyocera. IRIDIUM services are also used on a subscription basis by aircraft and vessels, thanks to dedicated transceivers units. Moreover, short IRIDIUM-compliant transceivers are also available and currently used within \ac{IoT} products to provide backhaul connections~\cite{iridium_iot}.

Each IRIDIUM satellite is characterized by a three-antenna array, forming a honeycomb pattern of 48 beams on the ground. Therefore, a generic IRIDIUM user might experience two types of hand-off, i.e., either from one beam to another beam (intra-satellite) or from one satellite to another one (inter-satellite). 
In general, when the user is served by a beam at the edge of the honeycomb, the system manages the hand-off to a neighbor satellite.

The channels provided by the system can be divided into two categories, i.e., \emph{system overhead channels} and \emph{bearer service channels}~\cite{icao}. In this paper, we are interested in one specific \emph{system overhead} channel, i.e., the \acl{IRA} (IRA) broadcast channel. This channel operates at 
$1,626.27$~MHz, and it is an unencrypted downlink-only channel used to send \emph{\ac{IRA}} messages to individual subscriber units---to be used for handover operations. \ac{IRA} messages contain the following plain-text information: satellite ID, beam ID, location at ground (in terms of longitude and latitude, as computed by the satellite based on a proprietary algorithm), altitude information (typically around $800$~km), and other information useful for handover purposes, such as the Temporary Mobile Subscriber Identity (TMSI) of the end-user equipment performing the handover. Note that, being an unencrypted channel, the IRA broadcast channel can be received by using a general-purpose \acl{SDR} (\ac{SDR}) and a generic antenna.

The position information computed by the IRIDIUM satellites and emitted through the broadcast IRA channel is used in this paper to compute a location estimation at the receiver side. This information is used to cross-check the position estimated from the GNSS constellation. 

\section{Adversary and System Model}
\label{sec:adversary_model}

Our adversary model consists of a malicious entity, able to generate fake GNSS signals by resorting to the combination of an \ac{SDR}, an antenna, and a GNSS spoofing software tool, such as~\cite{gpssdrsim}. 

\begin{figure}
\includegraphics[width=\columnwidth]{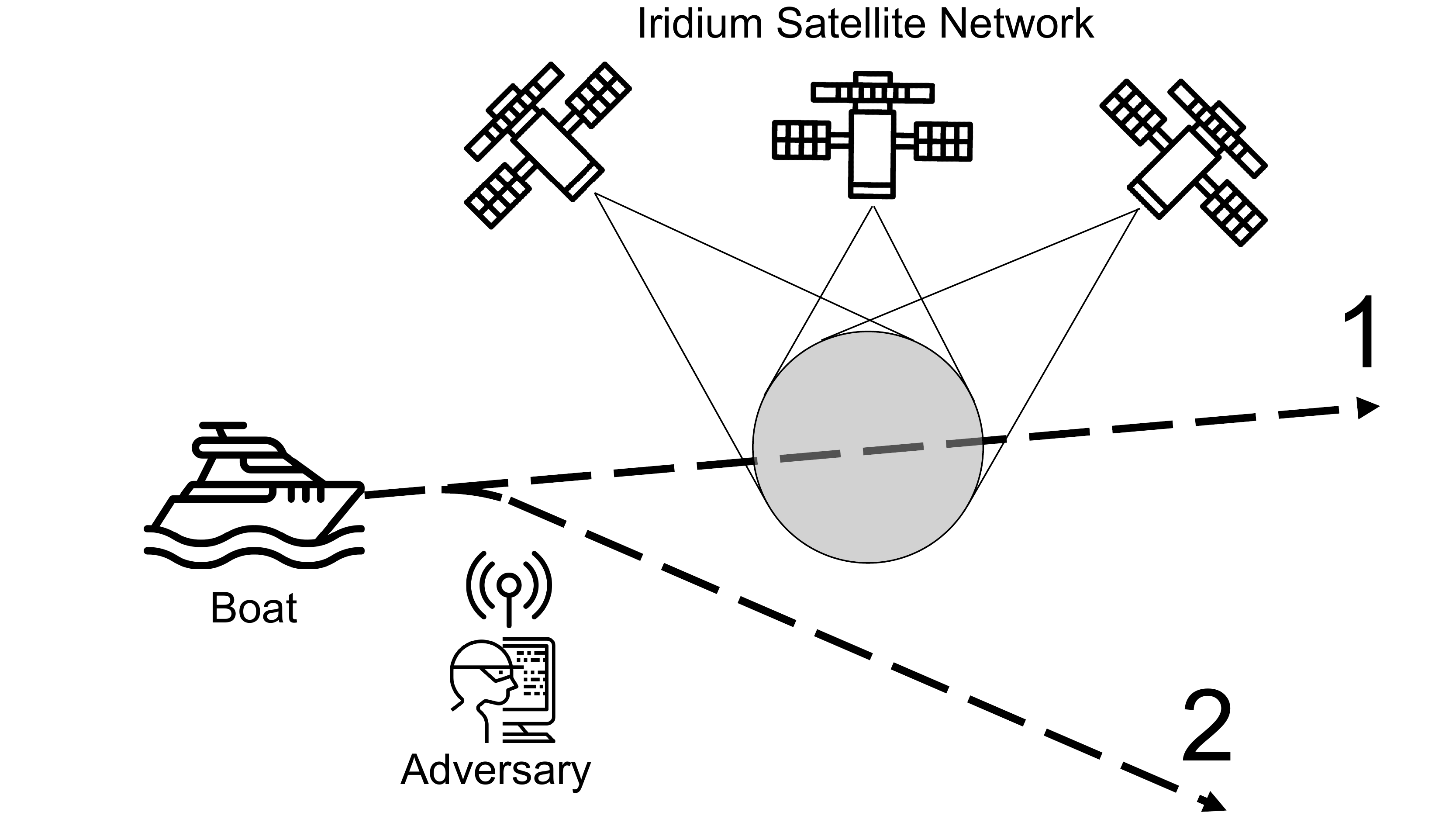}
\centering
\caption{Scenario and Adversary model: The target (boat) is re-routed by the adversary from path 1 to path 2, but our solution enables independent location verification by exploiting the IRIDIUM satellite network. Therefore, eventually the target will be able to detect the GNSS spoofing attack.}
\label{fig:adversary_model}
\end{figure}

We assume an isolated playground, not relying on any Internet connection and any other means (e.g., WiFi and Cellular Networks) to derive location information.
A typical scenario that fits the previous assumptions might be constituted by wide inhabited open areas, such as deserts, or open seas (oceans), where the target devices cannot receive any additional information to verify their actual GNSS location. 
Figure~\ref{fig:adversary_model} shows the system model considered throughout this paper. As a reference scenario, we consider a boat relying on GNSS information for its navigation. The adversary is willing to de-tour the boat by performing a GNSS spoofing attack; in particular, the adversary wants to re-route the boat from path $1$ to path $2$ in Fig.~\ref{fig:adversary_model}. Our solution exploits the IRIDIUM satellite network to independently verify the actual position of the boat, and therefore, raising an alarm when a de-tour is detected. 

It is worth noting that we do not assume any active IRIDIUM subscription. Thus, the IRIDIUM position-providing service  \ac{STL}
cannot be used. Indeed, purchasing unlimited subscription plans to IRIDIUM services can be particularly expensive for shipowners, and it could also require the hiring of dedicated teams to perform the deployment and maintenance in all the vessels. On the contrary, our solution exploits opportunistic IRIDIUM messages; in particular, the \ac{IRA} messages delivered in clear-text on the \emph{Ring Alert Broadcast Channel} (at the frequency $1,626.27$~MHz). 
Indeed, being unencrypted these messages can be received with a generic low-cost SDR.

In our system model, as depicted by the gray circle in Fig.~\ref{fig:adversary_model}, the boat continuously performs a loose localization process, and therefore, it checks if the current, locally computed, GNSS position is far away from the position provided by our solution exploiting the IRIDIUM constellation. 
In the remainder of this paper, we will estimate the configuration parameters and evaluate the thresholds to provide timely, independent, and effective verification of the GNSS localization.

\section{Measurement setup}
\label{sec:measurement_setup}

\begin{figure}
\includegraphics[width=0.8\columnwidth]{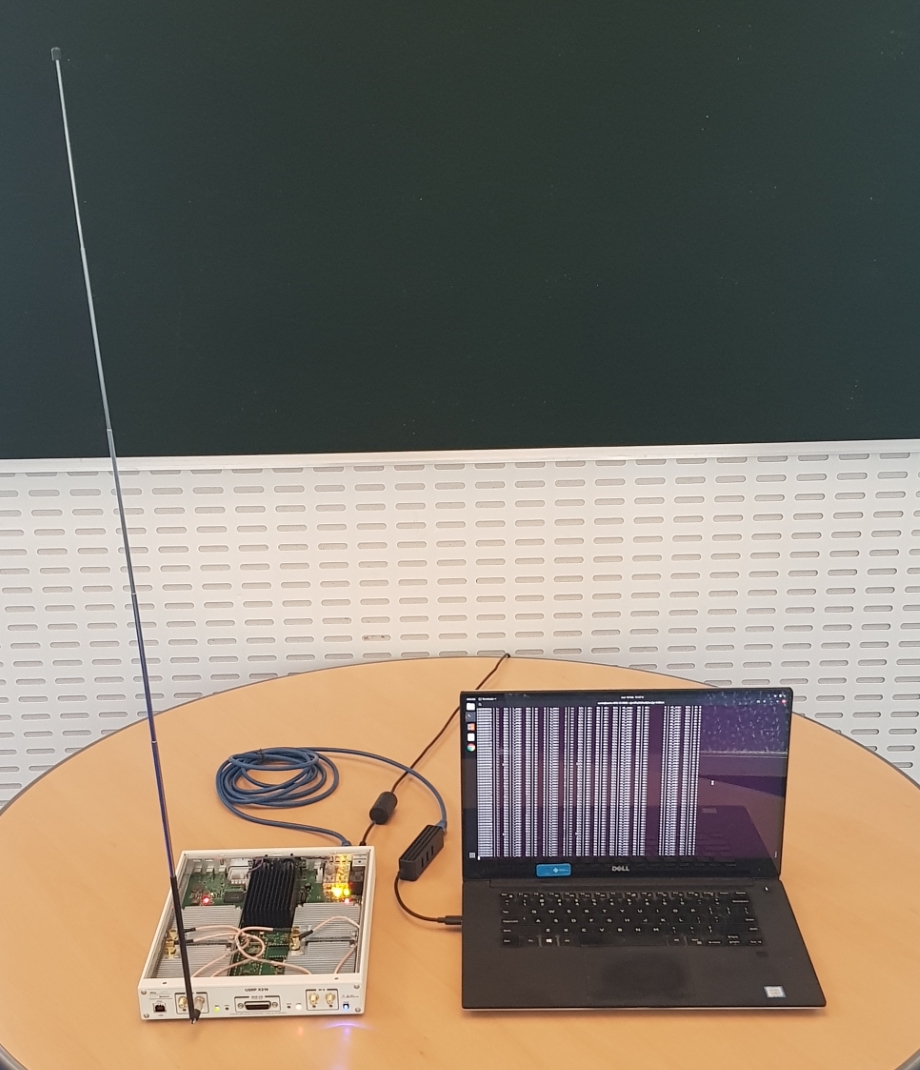}
\centering
\caption{Our receiver setup to collect the \ac{IRA} messages: a Dell XPS15 9560 laptop and a USRP X310.}
\label{fig:setup}
\end{figure}

Figure~\ref{fig:setup} shows the setup we adopted to receive the messages from the Iridium constellation. The hardware is mainly constituted by a Software Defined Radio USRP X310, a telescopic \emph{stilo} antenna, and a Laptop Dell XPS15 9560, equipped with 32GB of RAM and 8 Intel Core i7700HQ processors running at 2.80 GHz. As for the software, we adopted the GNURadio development toolkit and the related Iridium module \cite{iridiumgr}. The output is subsequently parsed to generate meaningful data analytics, available for downloading at \cite{crilab}.

Our measurement campaign has been carried out in very harsh conditions:  from a balcony outside our office. This sub-optimal setup significantly affects the number of received packets, being the antenna very close to the wall of the building. Moreover, the stylo antenna is recommended for operation from 75MHz to 1GHz, being therefore not optimal for the reception of the IRIDIUM signals at 1626.270833 MHz. While our setup still guarantees the collection of a reasonable amount of messages in a long term measurement, due to a high packet loss, it does not provide the best performance for a live-test of GPS spoofing detection. In later sections, we provide more details on statistics related to packet loss and how the later one affects our solution. The good results achieved in such harsh conditions undoubtedly show the viability of our approach.

We collected a total of about $1019$~hours of measurements consisting of $569,431$ samples. An excerpt from the dataset is reported in Table~\ref{table:excerpt}. For each data sample, we have the time (seconds and milliseconds), the satellite ID, the beam ID, and both latitude and longitude of the satellite/beam at the ground. Specifically, when the Beam ID is equal to zero, the position (latitude/longitude) refers to the position of the satellite at the ground, e.g., satellite 115 in the example of Table~\ref{table:excerpt}. When the Beam ID is different than zero, latitude and longitude refer to the position of the beam at the ground, e.g., the second sample is related to beam 44 belonging to satellite 115 having a position at the ground of 
$\langle +23.06, +49.81 \rangle$.
In all our measurements we counted a total of 66 satellite IDs, i.e., \{2, 3, 4, 5, 6, 7, 8, 9, 13, 16, 17, 18, 22, 23, 24, 25, 26, 28, 29, 30, 33, 36, 38, 39, 40, 42, 43, 44, 46, 48, 49, 50, 51, 57, 65, 67, 68, 69, 71, 72, 73, 74, 77, 78, 79, 81, 82, 85, 87, 88, 89, 90, 92, 93, 94, 96, 99, 103, 104, 107, 109, 110, 111, 112, 114,  115\}, coinciding with the number of active IRIDIUM satellites reported by the service provider, and the full-beam constellation equal to 48 beam IDs, i.e., $\{1, \ldots, 48\}$.

\begin{table}[htbp]
\footnotesize
\caption{Excerpt of the collected dataset.}
\begin{tabular}{cccccc}
\textbf{Time (s)} & \textbf{Time (ms)} & \textbf{\begin{tabular}[c]{@{}c@{}}Satellite \\ ID\end{tabular}} & \textbf{\begin{tabular}[c]{@{}c@{}}Beam \\ ID\end{tabular}} & \textbf{Latitude} & \textbf{Longitude} \\ \hline
1580712040 & 000000739 & 115 & 0  & +29.81            & +046.10            \\
1580712040 & 000004519 & 115 & 44 & +23.06            & +049.81            \\
1580712040 & 000005059 & 115 & 46 & +25.95            & +051.69            \\
1580712040 & 000005599 & 115 & 47 & +26.94            & +047.71            \\
1580712040 & 000008839 & 115 & 0  & +30.29            & +046.13            \\
1580712040 & 000013159 & 115 & 44 & +23.56            & +049.80            \\
1580712040 & 000013699 & 115 & 46 & +26.46            & +051.72           
\end{tabular}
\label{table:excerpt}
\end{table}


\section{IRIDIUM Data Analytics}
\label{sec:iridium_data_analytics}

In this section, we provide a thorough analysis of the network information that can be extracted from \ac{IRA} messages.

\begin{figure}
\includegraphics[width=\columnwidth]{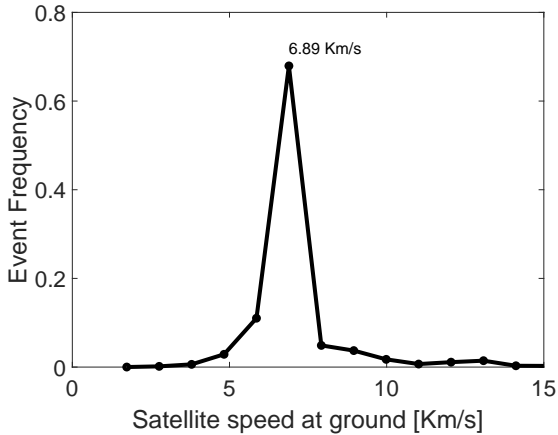}
\centering
\caption{Satellite speed at ground: probability distribution function associated with satellite speeds at ground and most likely speed (6.89Km/s).}
\label{fig:satellite_speed}
\end{figure}

{\bf Satellites speed.} Figure~\ref{fig:satellite_speed} shows the probability distribution function associated with the satellites speed at the ground. We took into account all the satellites IDs collected during our measurement campaign, by considering their positions and the absolute time associated with the reception of the packets. 
The satellite speed $v$ has been computed as the ratio of the difference between two consecutive positions, i.e., $\delta$, and the elapsed time, i.e., $\epsilon$, yielding $v = \frac{\delta}{\epsilon}$. We observe that the most likely speed is about $6.89$ Km/s, consistent with the approximate speed of a generic LEO satellite~\cite{ganz94}. 

We also notice the presence of many (unlikely) high-speed samples ($v>10$ Km/s). These inconsistent speeds are experienced when the time difference between two consecutive messages is particularly high, likely due to the packet loss. In particular, packet loss, especially when experienced in bursts, accumulates a lot of errors as for time and space, leading to inconsistent high speeds. Since we are not aware of the actual accuracy of beam and satellite coordinates at the ground, we guess that large periods between packets (due to packet loss) accumulate the aforementioned errors. However, we notice that this phenomenon does not affect the quality of our solution.

\begin{figure}
\includegraphics[width=\columnwidth]{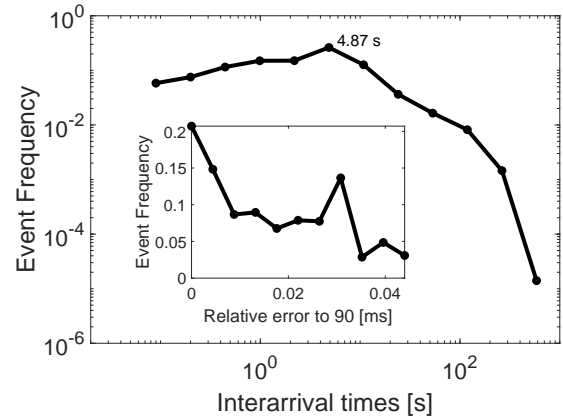}
\centering
\caption{\ac{IRA} interarrival times: probability distribution function associated to \ac{IRA} interarrival times and their maximum likelihood at 4.87 seconds. The inset figure represents the relative error respect to 90ms representing the actual rate at the transmitting side (for different beams).}
\label{fig:timings}
\end{figure}

{\bf \ac{IRA} message rate.} All our analysis is based on the reception of \ac{IRA} messages. Figure~\ref{fig:timings} shows the probability distribution function associated with the interarrival times of the \ac{IRA} messages. The inset figure shows the relative error to the expected interarrival time of 90 milliseconds, as reported by related work on IRIDIUM~\cite{leng}\cite{Tan2019_access}. 
Indeed, each beam transmits an \ac{IRA} message every $4.32$~seconds. Being the number of beams 48, a receiver will receive a generic \ac{IRA} message every 90ms ( $4320/48$). We observe also that the link is characterized by an extremely high packet error rate. 
Indeed, the likelihood for the interarrival time is about 4.87 seconds, i.e., one message is received every 66 lost ones. Conversely, the jitter is small; indeed, the computational overhead for message decoding is negligible and the vast majority of the messages experience an interarrival time that is multiple of 90 ms.

\begin{figure}
\includegraphics[width=\columnwidth]{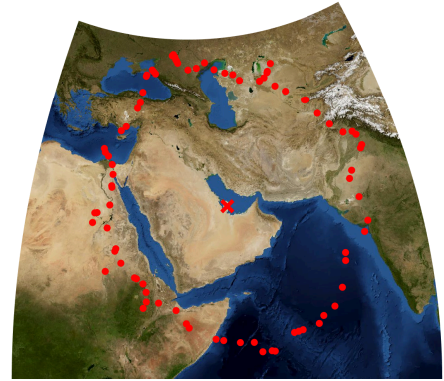}
\centering
\caption{Maximum satellite distances (at ground) from the receiving source.}
\label{fig:coverage}
\end{figure}

{\bf Coverage.} To understand the maximum coverage range of an Iridium-compliant device, we consider the furthest satellite positions at the ground, independently of the time they have been collected. Figure~\ref{fig:coverage} shows the furthest satellite positions (at ground) from our location
represented by the red cross. Many factors can affect the reception of \ac{IRA} messages: mountains in the north-east and south-west significantly affect the signal propagation, especially when the satellite is far-away from the receiver, at the minimal altitude over the horizon, or simply the presence of buildings. 
The overall area covered through our setup sums up to about $10^6  \cdot  8.3 km^{2}$, further demonstrating the capability of a very simple setup to cover even large areas.

\begin{figure}
\includegraphics[width=\columnwidth]{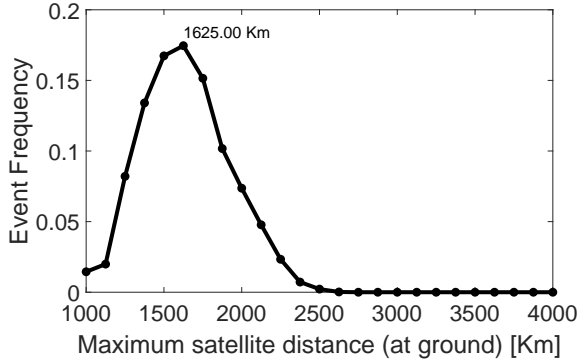}
\centering
\caption{Probability distribution function associated with the maximum satellite distances (at ground) from the receiving source. The maximum likelihood is $1,625~$Km.}
\label{fig:sat_distance}
\end{figure}

Figure~\ref{fig:sat_distance} shows the probability density function associated with the distance between the receiver and the satellite positions at the ground. As previously discussed, the transmission distance is significantly affected by the geography characterizing the neighborhood of the receiver. The likelihood of the maximum distance is $1,625$~Km. Note that a similar result can be extracted also by fitting the red dots in Fig.~\ref{fig:sat_distance}, using the Pratt method \cite{pratt}, obtaining a distance of about $2,346$~Km.

\begin{figure}
\includegraphics[width=\columnwidth]{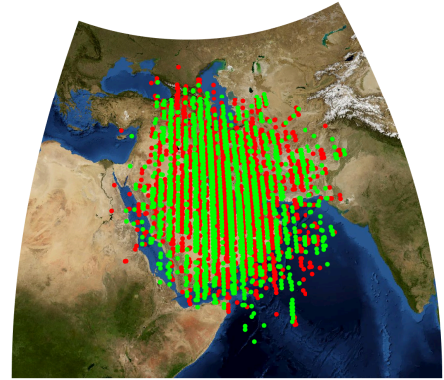}
\centering
\caption{Satellite 78 passes: data collected from Iridium satellite 78, where red and green dots represent upward and downward passes, respectively.}
\label{fig:satellite78}
\end{figure}

{\bf Satellite pass.} Let us consider a specific satellite, i.e., the Iridium satellite 78. Figure~\ref{fig:satellite78} shows all the passes extracted from all the measurements we collected about the aforementioned satellite. We adopted the red and green colors for the upward and downward directions, respectively. The overall coverage is consistent with the data we previously discussed for Fig.~\ref{fig:coverage}, while we observe a total of $59$ passes upward and $59$ downward in a period of time of about $1000$~hours.

\begin{figure}
\includegraphics[width=\columnwidth]{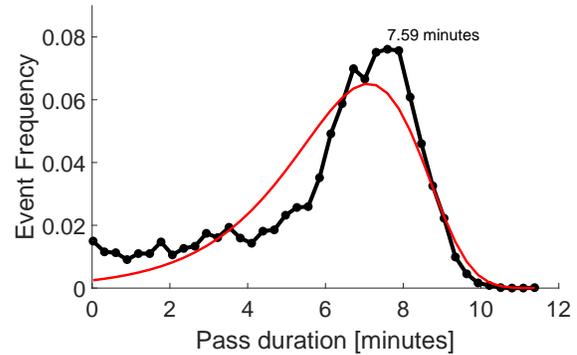}
\centering
\caption{Pass duration: probability distribution function associated with the duration of a satellite pass.}
\label{fig:pass_duration}
\end{figure}

The pass duration can be computed by considering the time between the instant when the satellite appears at the receiver and the last sample before disappearing. Figure~\ref{fig:pass_duration} shows the probability distribution function associated with the pass duration, considering all the satellites data collected from all our measurements. The maximum likelihood for the pass is about $7.59$ minutes. The distribution presents a heavy tail on the left side, i.e., pass duration less than 4 minutes: this is due to satellites crossing the coverage region far away from the receiver, and therefore, being characterized by short periods of intersection with the coverage area of the receiver. We searched for the best fit distribution and we found that to be the Extreme Value Distribution $evd(t)$, as depicted by Eq.~\ref{eq:passes}.

\begin{equation}
\label{eq:passes}
        evd(t) = \frac{1}{\sigma} \exp{\bigg(\frac{t-\mu}{\sigma}\bigg)} \exp{\bigg(-\exp{\bigg(\frac{t-\mu}{\sigma}\bigg)}\bigg)}, 
\end{equation}

where $\mu=7.28$ and $\sigma=1.67$ are the best-fit location parameter and the scale parameter of the distribution, respectively.

\begin{figure}
\includegraphics[width=\columnwidth]{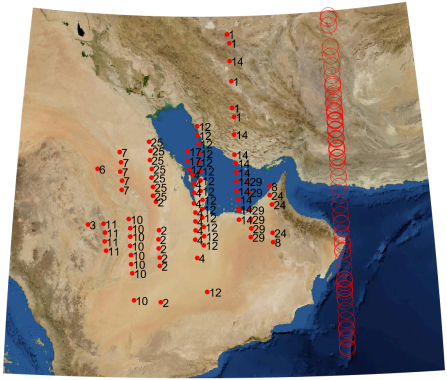}
\centering
\caption{Single satellite pass and beams: red circles represent the position of satellite 78 (at ground) while red dots show the beams' position at the ground. Beams IDs are reported accordingly to the information collected from the \ac{IRA} message.}
\label{fig:satellite78_beams}
\end{figure}

{\bf Satellite and beams.} Figure \ref{fig:satellite78_beams} shows an example pass (red circles) of satellite 78 and the beams (red dots) with the related beam IDs. We recall that both the beam positions at the ground and the beam IDs have been collected from the \ac{IRA} messages. Due to the previously mentioned packet loss rate, many beams are missing and only a subset of them are available at the receiver. In the aforementioned case study, we were able to retrieve 89 beam positions belonging to 15 different beams (for one satellite pass).

\begin{figure}
\includegraphics[width=\columnwidth,height=80mm]{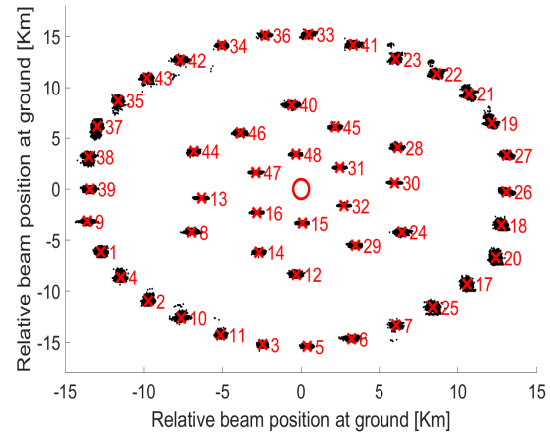}
\centering
\caption{Relative beam positions (at ground) respect to the satellite position (at ground). Each black dot represents a different measurement while crosses represent the centroid of each cloud associated to a different beam. Labels are consistent with the beam identifiers retrieved from the \ac{IRA} messages.}
\label{fig:beam_positions}
\end{figure}

{\bf Beam constellation.} We consider all the satellite beams collected from all our measurements and we computed their displacement to the satellite position. Black dots in Fig.~\ref{fig:beam_positions} show the beam constellation, while red crosses represent the centroid for each cloud of beams. The labels are consistent with the beam IDs retrieved from the \ac{IRA} messages. The beam constellation is constituted by 48 beams organized in three concentric circles with radii of about 3.36, 7.98, and 14.35 Km, respectively.

Finally, we observe that, to represent the beams in a well-organized constellation, we had to detect, distinguish, and subsequently compensate, satellites moving north  (upward) from the ones moving south (downward). Indeed, as previously discussed, data collected from beams might seem inconsistent, but actually, it is not, when making the following considerations:

\begin{itemize}
    \item \emph{Same beams belonging to the same satellite moving on different directions.} The same satellite can be observed going north, and subsequently south, after approximately 650 minutes. Indeed, due to its inclined orbit, the satellite will be back to the original position with the opposite transit direction. 
    \item \emph{Same beams belonging to different satellites moving on different directions.} Each location experiences different satellites going to different directions, either north or south.
\end{itemize}

For both the above configurations, the beam IDs should be mirrored to the x-axis. In particular, the beam constellation reported in Fig.~\ref{fig:beam_positions} is consistent with satellites going upward, i.e., from the south pole to the north pole. 

\section{GPS Spoofing Detection}
\label{sec:position_error_estimation}

In this section, we provide the details of our GNSS spoofing detection technique, aimed at detecting GNSS spoofing attacks using the IRIDIUM satellite constellation. Section~\ref{sec:pos_error} introduces the error in the position estimation via the \ac{IRA} messages, Section~\ref{sec:solution} presents the details of our spoofing detection solution, Section~\ref{sec:performance} shows the performance of the described technique, while Section~\ref{sec:spoofing} discusses the feasibility of a combined spoofing attack involving both the GNSS system and the IRIDIUM constellation.

\subsection{Position error estimation}
\label{sec:pos_error}

In this subsection, we estimate the error between the actual receiver position (taken from the GPS) and the one computed from the information extracted from the \ac{IRA} messages. The intuition behind our solution consists of computing the mean of all the latitude and longitude coordinates (centroid) collected from the beams. In general, the precision of the position estimation will be affected by the number of collected \ac{IRA} messages, and therefore, the time required to receive such a number of messages (waiting time). 

\begin{figure}
\includegraphics[width=\columnwidth]{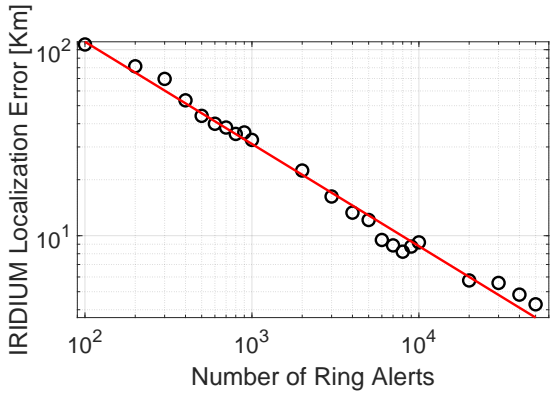}
\centering
\caption{Localization error as a function of the number of received \ac{IRA} messages. }
\label{fig:position_error}
\end{figure}

Figure~\ref{fig:position_error} shows the error between the receiver position and the estimated location, as a function of the number of collected \ac{IRA} messages. We observe that an error of about 10Km can be reached by collecting about 6,100 \ac{IRA} messages. 

Interestingly, the relationship between the localization error $L_{err}$ and the number of \ac{IRA} messages $n$ is characterized by a power-law, as depicted by the red line in Fig.~\ref{fig:position_error}, that in turn, is represented by Eq.~\ref{eq:loc_err}:
\begin{equation}
\label{eq:loc_err}
        L_{err} = n^m \cdot 10^q, 
\end{equation}
where $m=-0.5974$ and $q=3.2826$.

\begin{figure}
\includegraphics[width=\columnwidth]{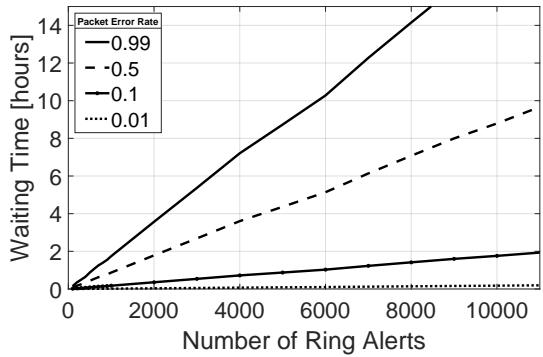}
\centering
\caption{Waiting Time (delay) to collect the \ac{IRA} messages as a function of the packet error rate.}
\label{fig:time_ring}
\end{figure}

Figure~\ref{fig:time_ring} shows the relation between the number of \ac{IRA} messages and the time to collect them, given different packet error rates, i.e., $\{0.01, 0.1, 0.5, 0.99\}$. 
In our scenario, given the adopted equipment and setup, we experienced a very high packet error rate (CFR. Fig.~\ref{fig:timings}), summing up to about 1 message received every 66, i.e., a packet error just slightly less than $0.99$.
In better conditions, lower packet error rates can be easily achieved, leading to a significantly higher amount of collected packets per unit of time. As depicted in Fig.~\ref{fig:time_ring}, in the worst-case scenario---our deployment, $6100$ packets, guaranteeing a localization error of about 10Km, require about 10 hours. Assuming lower packet error rates and good receiving conditions (such as in a desert or ocean), the same amount of packets can be collected in 1 hour or even in 6 minutes, assuming a packet error rate of either 10\% or 1\%, respectively.

\subsection{Exploiting IRA messages to estimate receiver location}
\label{sec:solution}

Our solution involves a multi-stage algorithm described in the following.

\begin{enumerate}
    \item {\bf Collect data.} The user continuously collects \ac{IRA} messages from the IRIDIUM constellation network, as described in the previous sections.
    \item {\bf Compensate movement.} Given the movement of the receiver during the message collection, the acquired data should be compensated as a function of the movement. Therefore, assuming the acquisition process starts at $t_0$, and the user is moving with speed $v$, the collected information from the \ac{IRA} messages should be compensated as reported by Eq.~\ref{eq:comp}.
    \begin{eqnarray}
    \label{eq:comp}
        \Delta x & = & x(t_0) + \cos(v \cdot \Delta t), \nonumber\\
        \Delta y & = & y(t_0) + \sin(v \cdot \Delta t), 
    \end{eqnarray}

    where $\Delta t$ is the time between two consecutive IRIDIUM packets, while $x$ and $y$ are latitude and longitude, respectively.
    
    \item {\bf Compute location $I_{pos}$.} The position of the receiver is estimated as the centroid computed over the  beams' positions at ground, yielding Eq.~\ref{eq:centroid}.
    \begin{equation}
    \label{eq:centroid}
        I_{pos} = \bigg(\frac{1}{N} \sum_{i=1}^{N} x_i,\quad \frac{1}{N} \sum_{i=1}^{N} y_i\bigg),
    \end{equation}
    where $(x_i, y_i)$ are the latitude and longitude of the position at ground for beam $i$ , while $N$ is the number of beams taken into account for the computation (cfr.  Fig.~\ref{fig:position_error}).
    
    \item {\bf Compare $I_{pos}$ with $G_{pos}$.} The estimated position $I_{pos}$ from the IRIDIUM satellite network is compared with that one received from the GNSS system $G_{pos}$. Depending on different parameters, that will be discussed later on, an alarm is raised if the difference is greater than a predetermined threshold $thr$, i.e., $\mid I_{pos} - G_{pos} \mid > thr$.
\end{enumerate}

\subsection{Performance}
\label{sec:performance}

Our solution is particularly suitable for scenarios where crowd-sourcing is not possible, i.e., there are no other independent sources in the neighborhood to verify the current GNSS localization. A few examples of such scenarios are deserts, forests, remote rural areas, and open seas. We choose the latter one as our reference scenario, and without loss of generality, we consider a ship moving through a pre-defined path according to different speeds. According to the specific cruise speed, we provide different indicators to measure the performance of our solution.

In the following, we consider different classes of ships~\cite{marineinsight}, as depicted in Table~\ref{tab:shiptype}. Class S1, i.e., \emph{bulk carriers}, carry unpacked bulk cargo, class S2, i.e., \emph{Container Ships}, carry containers, S3, i.e., \emph{Oil and chemical tankers}, are designed specifically to carry oil and chemicals, S4, i.e., \emph{RORO vessels}, roll on roll off ships are special types of vessels used for the transportation of automobile vehicles, and finally, S5, i.e., \emph{Cruise Ships}, are luxury vessels used to take passengers on pleasure journey.

\begin{table}[]
\caption{Ship type and speeds.\label{tab:shiptype}}
\begin{tabular}{llll}
                        & \textbf{Ship Type}       & \textbf{\begin{tabular}[c]{@{}l@{}}Speed\\ {[}knots{]}\end{tabular}} & \textbf{\begin{tabular}[c]{@{}l@{}}Speed \\ {[}km/h{]}\end{tabular}} \\ \hline
\multicolumn{1}{l|}{S1} & Bulk Carriers            & 13-15                                                                & 24-28                                                                \\
\multicolumn{1}{l|}{S2} & Container Ships          & 16-24                                                                & 30-44                                                                \\
\multicolumn{1}{l|}{S3} & Oil and chemical tankers & 13-17                                                                & 24-31                                                                \\
\multicolumn{1}{l|}{S4} & RORO vessels             & 16-22                                                                & 30-41                                                                \\
\multicolumn{1}{l|}{S5} & Cruise Ships             & 20-25                                                                & 37-46  
\end{tabular}
\end{table}

\begin{figure}
    \includegraphics[width=\columnwidth]{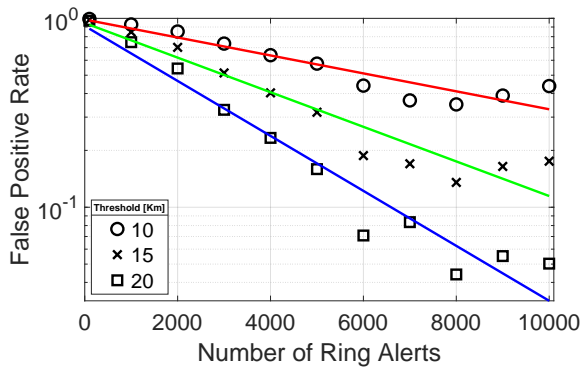}
    \centering
    \caption{False positive events as a function of the number of collected \ac{IRA} messages.}
    \label{fig:false_positive}
\end{figure}

\begin{table*}[h]
\centering
\caption{Overview of related work on GNSS spoofing detection and related system requirements. \label{tab:related}}
\begin{tabular}{|P{2cm}|P{2.2cm}| P{2.2cm}| P{2.cm}| P{2.cm}| P{2.cm}|}
\hline
 \textbf{Ref.} &  \textbf{No need for Auxiliary Ad-Hoc Infrastructure} & \textbf{No need for  Auxiliary In-land Infrastructure} & \textbf{No need for  PHY-layer Information} & \textbf{No need for  Dedicated Hardware} & \textbf{No need for Multiple Antennas} \\
\hline
 \cite{oligeri2019_wisec}     & \cmark & \xmark     & \cmark & \cmark & \cmark \\ \hline
 \cite{jansen2016_acsac}      & \cmark & \cmark & \xmark     & \xmark     & \xmark     \\ \hline
 \cite{milaat2018_commlett}   & \cmark & \cmark & \xmark     & \cmark & \xmark     \\ \hline
 \cite{zhang2012_milcom}      & \cmark & \cmark & \xmark     & \xmark     & \xmark     \\ \hline
 \cite{sciancalepore2018_cns} & \xmark     & \xmark     & \xmark     & \cmark & \xmark     \\ \hline
 \cite{Jansen2018}            & \xmark     & \xmark     & \xmark     & \cmark & \xmark     \\ \hline
 \cite{qiao2017_cis}          & \cmark & \cmark & \cmark & \xmark     & \cmark \\ \hline
 \cite{heng2015_tis}          & \xmark     & \xmark     & \xmark     & \cmark & \xmark     \\ \hline
 \cite{magiera2015_jart}      & \cmark & \cmark & \xmark     & \xmark     & \cmark \\ \hline
 Our Proposal                 & \cmark     & \cmark     & \cmark     & \cmark     & \cmark     \\ \hline
\end{tabular}
\hfill
\end{table*}

As previously discussed, the accuracy of the estimated location $I_{pos}$ depends on the number of collected messages. Indeed, recalling Fig.~\ref{fig:time_ring}, the number of collected messages is correlated to the time dedicated to collect them. 
Considering a location estimation error of about $10$~Km, meaning $6,100$ \ac{IRA} messages to be collected in about 6 minutes (packet loss equal to 1\%), the fastest ship, i.e., S5, would have moved off about 4.1Km (worst case). In the best-case scenario, the slowest ship, i.e., either S1 or S3, would have traveled less than 3.1 Km in the same amount of time (6 minutes). This means that, for all the considered cases, the movement performed by the ship is in the same order of the localization error, and therefore, our solution can be effectively adopted to detect GPS spoofing attacks.

\begin{definition}
    Given a threshold $thr$, we define a \emph{False Positive} event when no GNSS spoofing attack is present, and a position $I_{pos}$ computed exploiting the \ac{IRA} messages verifies $|I_{pos} - G_{pos}| > thr$ ---where $G_{pos}$ is the position computed from the GNSS constellation.
\end{definition}

Figure~\ref{fig:false_positive} shows the false positive rate as a function of the collected \ac{IRA} messages, considering three different thresholds, i.e., $thr \in \{10, 15, 20\}$ Km, for a number of \ac{IRA} messages varying from 10 to 10,000. We want to stress that we are considering \emph{single} false-positive events, while a more in-depth analysis considering the temporal distribution of false-positive events (bursts of consecutive events) might significantly improve the robustness of our solution to false alarms---this task is left for future work. 

Finally, it is interesting to observe that there is a relationship between the false positive rate and the number of \ac{IRA} messages considered in the location estimation technique. Indeed, we fit the samples belonging to the three different thresholds by using a linear regression model, yielding the solid lines in Fig.~\ref{fig:false_positive}, i.e., the red solid line for $thr=10$ Km, the green solid line for $thr=15$ Km, and finally, the blue solid line for $thr=20$km. The best-fit model associated with the aforementioned samples is represented by Eq.~\ref{eq:fp_fit}:
\begin{equation}
\label{eq:fp_fit}
    y = 10^{m \cdot n} \cdot 10^q,
\end{equation}

\noindent
where $n$ is the number of \ac{IRA} messages, $q \approx 0$, and $m=-4.6 \cdot 10^{-5}$, $m=-8.9 \cdot 10^{-5}$, $m=-1.4 \cdot 10^{-4}$ for $thr = 10$, $thr = 15$, and $thr = 20$, respectively.

Overall, Eq.~\ref{eq:loc_err} and Eq.~\ref{eq:fp_fit} allow an end-user deploying our  solution to have an immediate idea of the expected system performance. In particular, knowing the number of messages received in a given unit of time, the cited equations can be used to provide an estimation of the expected location accuracy and false-positives rate of the system---this latter one with the caveat that just a single value exceeding the threshold would trigger the alarm.

\subsection{Spoofing the IRIDIUM satellite constellation} 
\label{sec:spoofing}

With our solution in place, the only chance for the adversary to fool the receiver would be to perform a combined attack on both the GNSS and IRIDIUM platforms.
In the following, we discuss the feasibility of spoofing IRIDIUM signals. 
As detailed in the following, it runs out that, given the nature of our solution, spoofing IRIDIUM signals to subvert our localization system is more challenging compared to all the GNSS ones.

{\bf Time.} Our solution requires a not-negligible  amount of time (depending on the packet loss) 
to locate the vessel with a small precision error (less than 10Km). While GNSS requires seconds to be spoofed, to affect our IRIDIUM-based localization system, the adversary has to deploy a transmitter that should keep transmitting for either minutes or hours, thus becoming easier to detect, identify, and locate.

{\bf Signal strength.} The IRIDIUM satellite infrastructure is significantly different from the typical GNSS constellations. GNSS satellites are far away from the ground (in a MEO orbit, at about $20,000$~Km), while the Iridium constellation is composed of LEO satellites, orbiting at about $800$~Km. Therefore, the received signal strength on the ground is higher than the GNSS ones~\cite{tan2020_electronics}, and hence the adversary requires higher transmission power to overcome the legitimate signal.

{\bf Complexity.} The adversary should mimic a consistent array of beams moving coherently all together, at a consistent speed, 
with consistent interarrival times, and eventually, with compatible passes duration. We consider this event very unlikely, especially considering the time factor previously described. 

\section{System-Level Comparison with Other Solutions}
\label{sec:comparison}

To provide further insights, Table~\ref{tab:related} wraps up an overview of the system requirements for both the works previously discussed in Section~\ref{sec:related} and our proposal. We notice that most of the already proposed solutions rely on auxiliary network infrastructures. These auxiliary infrastructures can be set up ad-hoc for the spoofing detection task (such as in~\cite{sciancalepore2018_cns},~\cite{Jansen2018} and~\cite{heng2015_tis}), or they can be already in place for other purposes (e.g., the mobile cellular network used in~\cite{oligeri2019_wisec} and the avionics network in~\cite{Jansen2018}). 
However, these infrastructures are generally not available in remote or offshore areas, such as deserts and open seas. Other proposals, such as \cite{jansen2016_acsac},~\cite{milaat2018_commlett},~\cite{zhang2012_milcom}, and~\cite{heng2015_tis}, rely on multi-antenna detection schemes, requiring a dedicated setup and the hard swap of existing GNSS receivers. 
Moreover, schemes such as~\cite{magiera2015_jart} require access to physical layer information of the GNSS signal, and therefore, requiring dedicated equipment to be deployed, they come with consequently high costs.

Overall, the comparison shows that our proposal leveraging opportunistic IRIDIUM signals is the only one that does not require any auxiliary ad-hoc infrastructure, it does not leverage any physical layer information and, finally, it does not require the deployment of either multiple antennas or dedicated hardware. Therefore, our solution is ideal when a specific target receiver is meant to be operated in remote areas (desert or open seas), and whose hardware cannot be modified after deployment.

\section{Conclusion and Future Work}
\label{sec:conclusion}

In this paper, we have provided two major contributions: we have reverse-engineered several system features of the IRIDIUM satellite constellation, and we have proposed a new solution to detect GNSS spoofing attacks--- combining the cited features with the publicly available \acl{IRA} messages. We have provided an efficient and effective location verification algorithm, independent from GNSS technology, 
that can be used to verify the actual position computed from standard GNSS technologies. Overall, our solution is worldwide-available, cheap, non-invasive, and it does not require either information from the radio physical-layer or multiple antennas. Therefore, it is particularly suitable for unattended scenarios, where there are no other entities that can help to verify the actual position of the receiver. These scenarios include deserts, poles, and open seas, to name a few. The data have been obtained thanks to an extensive measurement campaign lasting over $1000$ hours, and they have been also made public to foster research in this challenging field.

Our future work will focus on the characterization of the power-law obeying phenomena highlighted by our experimental campaign, to improve the accuracy of the location verification scheme, as well as to analyze the time-correlation among false-positive events to declare a GNSS spoofing attack. We will also collect further measurements data, using newly acquired dedicated equipment, such as powerful IRIDIUM-compliant L-band antennas.

\section*{Acknowledgements}
This publication was partially supported by awards NPRP-S-11-0109-180242, NPRP12S-0125-190013, and NPRP X-063-1-014 from the QNRF-Qatar National Research Fund, a member of The Qatar Foundation. The information and views set out in this publication are those of the authors and do not necessarily reflect the official opinion of the QNRF.

\bibliographystyle{ACM-Reference-Format}
\balance
\bibliography{biblio}

\end{document}